\documentclass[a4paper]{paper}

\pdfoutput=1
\usepackage{amssymb}
\usepackage{amsmath}
\usepackage[T1]{fontenc}
\usepackage{lmodern}
\usepackage{graphicx}
\usepackage{geometry}

\title{Fraud detection with statistics: A comment on \emph{Evidential Value in ANOVA-Regression Results in Scientific Integrity Studies} (Klaassen, 2015).}

\author{Hannes Matuschek}
\institution{
 Focus Area for Dynamics of Complex Systems and
 Department of Psychology,\\
 University of Potsdam,
 Karl-Liebknecht-Str. 24,
 D-14476 Potsdam, Germany,\\
 hannes.matuschek@uni-potsdam.de}

\begin{document}

\maketitle

\begin{abstract}
Klaassen in \cite{Klaassen2015} proposed a method for the detection of data manipulation given the means and standard deviations for the cells of a oneway ANOVA design. This comment critically reviews this method. In addition, inspired by this analysis, an alternative approach to test sample correlations over several experiments is derived. The results are in close agreement with the initial analysis reported by an anonymous whistleblower \cite{Anonymous2012}. Importantly, the statistic requires several similar experiments; a test for correlations between 3 sample means based on a  single experiment must be considered as unreliable.
\end{abstract}

\section{Introduction}
An analysis of means and standard deviations \cite{Peeters2015}, culled from a series of scientific publications, led to a request for retraction of a subset of the papers \cite{UvA2015}. The analysis was based on a method reported in Klaassen \cite{Klaassen2015} aimed at detecting a type of data manipulation that causes correlations between condition means of samples that are assumed to  be independent. Specifically, given a one-way balanced ANOVA design with 3 conditions, $X_{i},i=1,...,3$, the means obtained by averaging over the scores of $n$ different subjects in each condition, are samples of a 3-dimensional normal distribution

\begin{equation}\label{eq:lik}
\left(\begin{array}{c} X_{1}\\ X_{2}\\ X_{3} \end{array}\right)
\sim \mathcal{N}\left(\left(\begin{array}{c} \mu_{1}\\ \mu_{2}\\ \mu_{3} \end{array}\right) , n^{-1}\left(\begin{array}{ccc}
 \sigma_{1}^{2} & \sigma_{1}\sigma_{2}\rho_{1} & \sigma_{1}\sigma_{3}\rho_{2}\\
 \sigma_{1}\sigma_{2}\rho_{1} & \sigma_{3}^{2} & \sigma_{2}\sigma_{3}\rho_{3}\\
 \sigma_{1}\sigma_{3}\rho_{2} & \sigma_{2}\sigma_{3}\rho_{3} & \sigma_{3}^{2}
\end{array}\right)\right) ,
\end{equation}
where $\mu_{i}$ are the unknown \emph{true} expected values and $\sigma_{i}$ the unknown sample standard deviations of the scores under the respective conditions and $\rho_{i}$ their correlations. The ANOVA assumes independence between the samples of the conditions, such that $\rho_{i}=0$. Indeed, given only samples of $X_{i}$ and estimates of $\sigma_{i}$, the sample correlations $\rho_{i}$ are not directly accessible. 

\begin{figure} [ht!]
 \centering
 \includegraphics[width=0.75\textwidth]{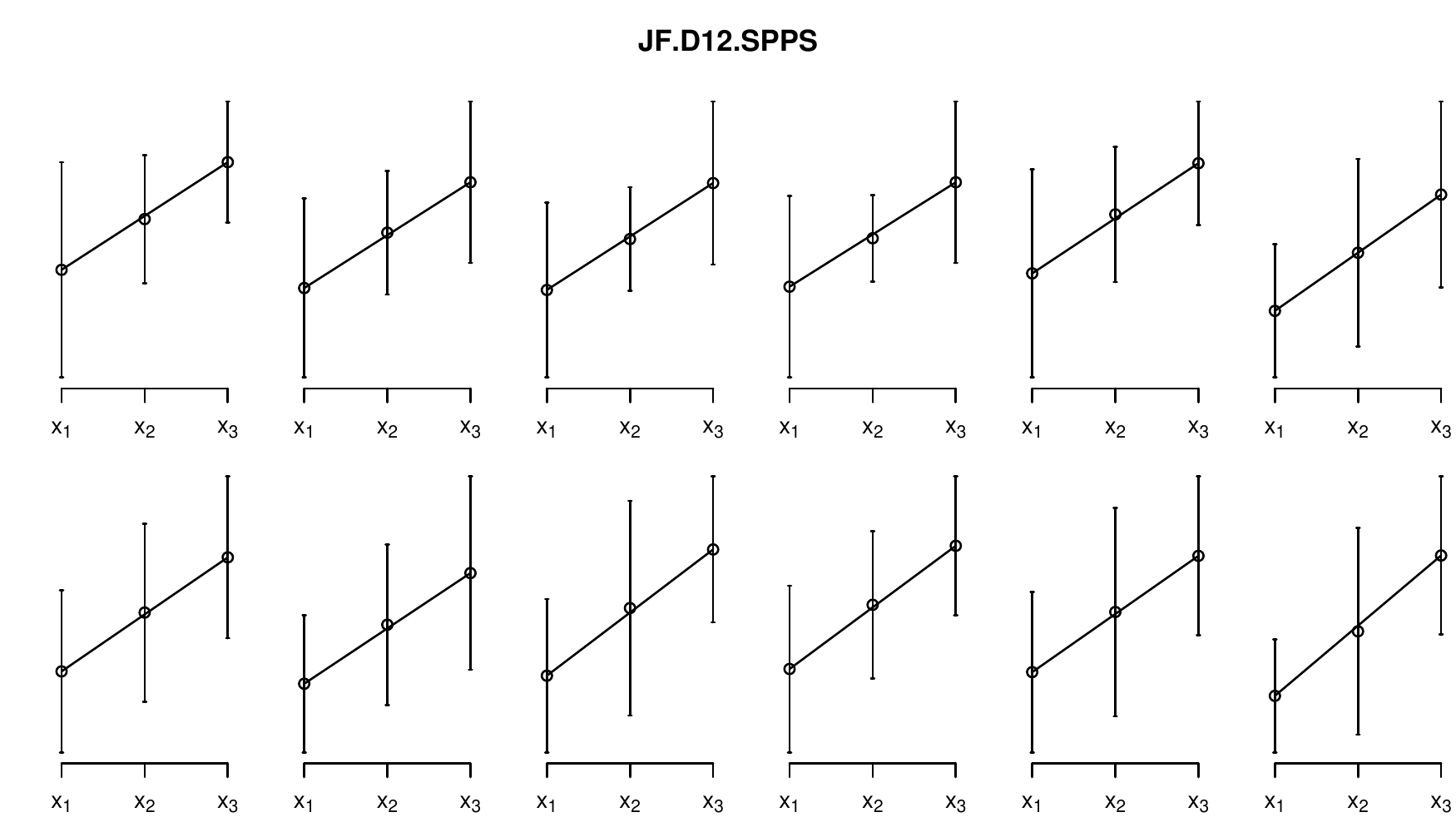}
 \caption{Condition means ($x_1$, $x_2$ and $x_3$) and standard deviations for the 12 experiments reported in \cite{Vision2014}. The condition means $x_1$ and $x_3$ have been connected by a line to visualize the deviance from a perfect linear behavior of the condition means. \label{fig:jfd12}}
\end{figure}

An anonymous whistleblower pointed out \cite{Anonymous2012}, that the results in the studies under suspicion (i.e \cite{Vision2014}, compare Figure \ref{fig:jfd12}), show a \emph{super linear} pattern which appears \emph{too good to be true}. Importantly, the authors of the original publications did not necessarily expect such patterns of equidistant means;  they expected an ordinal, not a linear relation between the three condition means. Nevertheless, the reanalyses were carried out under the assumption of an expected strict linear relation between means. The reason was that this strict assumption is conservative with respect to an inference of data manipulation\footnote{It is also not clear how a suitable test could be constructed for the assumption that the means are expected only in a monotonic, not necessarily equidistant order.}.

Under the assumption of a strictly linear  relationship between the group means, $\mu_i=\alpha+\beta\cdot i$, the scores can be described as $X_{i}=\alpha+\beta\cdot i+\epsilon_i$ which implies that  $0=E[Z]=E[X_{1}-2X_{2}+X_{3}]=\mu_1-2\mu_2+\mu_3$. This linear-combination of sample means $X_i$ yields a new random variable $Z$ with the (univariate) normal distribution $Z\sim\mathcal{N}(0,n^{-1}\sigma_{Z}^{2}(\vec{\sigma},\vec{\rho}))$. Where $\sigma_{Z}^{2}(\vec{\sigma},\vec{\rho}) = \sigma_{1}^{2} + 4\,\sigma_{2}^{2} + \sigma_{3}^{2} - 4\sigma_{1}\sigma_{2}\rho_{1} - 4\sigma_{2}\sigma_{3}\rho_{3} + 2\sigma_{1}\sigma_{3}\rho_{2}$. Note that the random variable $Z$ can be seen as the deviance
from the strictly linear behavior $\alpha+\beta\, i$. 

Introducing correlations between the samples increases or decreases the variance of $Z$. Klaassen \cite{Klaassen2015} assumes that a plausible  data manipulation (e.g., adjusting the mean of the middle sample towards the mean of the means of the lower and upper samples to achieve significant differences between the groups) leads to a decrease of the variance of $Z$, $\sigma_{Z}^{2}(\cdot,\cdot)$. Such a variance reduction may have gone unnoticed as \emph{humans tend to underestimate variance} in data. As mentioned above, the results under suspicion show a \emph{super linear} behavior and hence a small variance in $Z$ which may not be expected given the group variances $\sigma_{i}^{2}$ under the assumption of independence.

Consequently, Klaassen \cite{Klaassen2015} used a simple likelihood-ratio test to decide whether there is evidence for data manipulation in terms of a \emph{evidential value} as
\[
V=\frac{\underset{\vec{\rho}\in\mathcal{F}}{\text{max}\,}f(z|\sigma_{z}(\vec{\sigma},\vec{\rho}))}{f(z|\sigma_{z}(\vec{\sigma},\vec{0}))}\,,
\]
comparing the maximum likelihood of all feasible vectors of correlations
$\vec{\rho}$ with the likelihood of $z$ under the assumption of
$\vec{\rho}=\vec{0}$, where 
\begin{equation*}
 \mathcal{F} = \left\{ \vec{\rho}:\rho_{i}\in(-1,1),\,\rho_{1}^{2}+\rho_{2}^{2}+\rho_{3}^{2}-2\rho_{1}\rho_{2}\rho_{3}<1,\right.
 \left. \,\sigma_{Z}(\vec{\sigma},\vec{\rho})\le\sigma_{Z}(\vec{\sigma},\vec{0})\right\} \,,
\end{equation*}
is the set of feasible correlation vectors, maintaining that the covariance
matrix (in eq. \ref{eq:lik}) remains positive definite and ensures
that $\sigma_{Z}(\vec{\sigma},\vec{\rho})\le\sigma_{Z}(\vec{\sigma},\vec{0})\,\forall\,\vec{\rho}\in\mathcal{F}$.
As the \emph{true} sample standard deviations $\vec{\sigma}$ are unknown, they
might be replaced by the reported ones $\vec{s}$, since
$\vec{s}\rightarrow\vec{\sigma}$ as $n\rightarrow\infty$.

\section{An asymptotic test statistic\label{sec:teststat}}
Without any knowledge of the test statistics, i.e. the distribution of $V$ under the null hypothesis $H_{0}$ (independent group means), it is not possible to interpret the value $V$ and hence to decide whether a certain value of $V$ does provide evidence for the presence of sample correlations. The estimates of the sample variances ($\vec{s}{}^{2}$) are themselves random variables with some unknown distribution. It is therefore rather unlikely to obtain a closed form expression for the test statistic, even under restrictive assumptions about the distribution of $\vec{s}$. 

Nevertheless, as proposed by Klaassen  \cite{Klaassen2015}, one may assume that asymptotically $\vec{s}\rightarrow\vec{\sigma}$ as $n\rightarrow\infty$. Then one can assume that the sample variances $\vec{\sigma}$ are fixed and known, allowing for the construction of an upper-bound asymptotic test statistic.

The likelihood to obtain a specific value $z$, given the sample variances $\vec{\sigma}^{2}$ and correlations $\vec{\rho}$ is
\[
f(Z=z|\sigma_{Z}(\vec{\sigma},\vec{\rho}))=\frac{\sqrt{n}}{\sqrt{2\pi}\sigma_{Z}(\vec{\sigma},\vec{\rho})}\exp\left\{ -\frac{n\, z^{2}}{2\,\sigma_{Z}^{2}(\vec{\sigma},\vec{\rho})}\right\} 
\]
and therefore 
\begin{equation}
V = \underset{\vec{\rho}\in\mathcal{F}}{\text{max}}\, \frac{\sigma_{Z}(\vec{\sigma},\vec{0})}{\sigma_{Z}(\vec{\sigma},\vec{\rho})} \exp\left\{ -\frac{n\, z^{2}}{2\,\sigma_{Z}^{2}(\vec{\sigma},\vec{\rho})} + \frac{n\, z^{2}}{2\,\sigma_{Z}^{2}(\vec{\sigma},\vec{0})} \right\} \,.
\end{equation}

Now, let $a=\frac{\sigma_{Z}(\vec{\sigma},\vec{\rho})}{\sigma_{Z}(\vec{\sigma},\vec{0})}$
be the relative standard deviation and $\sigma_{0}=\sigma_{Z}(\vec{\sigma},\vec{0})$
then
\[
V=\underset{a\in\mathcal{A}}{\text{\text{max}}}\, a^{-1}\,\exp\left\{ -\frac{n\, z^{2}}{2\, a^{2}\sigma_{0}^{2}}+\frac{n\, z^{2}}{2\sigma_{0}^{2}}\right\} \,.
\]

The feasible set of all $a$ values $\mathcal{A}$ is implicitly defined by the feasible set of correlations as 
\[
\mathcal{A}=\left\{ \frac{\sigma_{z}(\vec{\sigma},\vec{\rho})}{\sigma_{z}(\vec{\sigma},\vec{0})}:\:\vec{\rho}\in\mathcal{F}\right\} \,.
\]

From this it follows immediately that $\mathcal{A}\subseteq(0,1]$
as $\sigma_{Z}(\vec{\sigma},\vec{\rho})\leq\sigma_{Z}(\vec{\sigma},\vec{0})\,\forall\,\vec{\rho}\in\mathcal{F}$. 

Under a \emph{worst-case} scenario, one may assume $\mathcal{A}=(0,1]$. This implies that for every $a\in(0,1]$ it is possible to find a feasible correlation vector $\vec{\rho}\in\mathcal{F}$ such that $\sigma_{Z}(\vec{\text{\ensuremath{\sigma}}},\vec{\rho})=a\,\sigma_{0}$. Please note that this is not ensured in general. The \emph{worst-case} assumption, however, allows one to obtain upper-bounds for the distribution of $V$ under $H_{0}$ analytically by relaxing the constraints on $a$ implied by the feasibility constraints on $\vec{\rho}$.

Within this setting one gets
\[
V\le\hat{V}=\underset{a\in(0,1]}{\mbox{max}}a^{-1}\,\exp\left\{ -\frac{n\, z^{2}}{2a^{2}\sigma_{0}^{2}}+\frac{n\, z^{2}}{2\sigma_{0}^{2}}\right\} \,.
\]

With $\tilde{z}=\frac{\sqrt{n}z}{\sigma_{0}}$, the normalized $z$ with respect to the expected standard deviation under $H_{0}$
\[
\hat{V}=\underset{a\in(0,1]}{\mbox{max}}a^{-1}\,\exp\left\{ -\frac{\tilde{z}^{2}}{2a^{2}}+\frac{\tilde{z}^{2}}{2}\right\} \,.
\]

Straightforward computation reveals
\[
0=\partial_{a}\left(\log\left[a^{-1}\,\exp\left\{ -\frac{\tilde{z}^{2}}{2a^{2}}+\frac{\tilde{z}^{2}}{2}\right\} \right]\right)\quad\Rightarrow\quad a^{2}=\tilde{z}^{2}\,,
\]
and therefore 
\begin{equation}
\hat{V}=\begin{cases}
1 & :\,\left|\tilde{z}\right|>1\\
\left|\tilde{z}\right|^{-1}\exp\left\{ \frac{\tilde{z}^{2}-1}{2}\right\}  & :\,\text{else}\,.
\end{cases}
\end{equation}

Under the \emph{worst-case} scenario, an upper-bound evidential value  $\hat{V}\ge V$ can be computed directly without maximizing the likelihood-ratio numerically. This result was also found by Klaassen  (compare eq. 18 in \cite{Klaassen2015}).

Knowing that the maximum $\hat{V}$ is achieved at $\tilde{z}^{2}=a^{2}$ and therefore $\frac{nz^{2}}{\sigma_{0}^{2}} = \frac{\sigma_{Z}^{2}(\vec{\sigma},\vec{\rho})}{\sigma_{0}^{2}}$, one may conclude that the likelihood-ratio test compares the expected variance $\sigma_{0}^2$ under $H_{0}$ with a variance estimated from a single sample. Such a variance estimate is known to be unreliable and therefore the evidential value for a single experiment must be unreliable, too. This issue is discussed in detail in the next section.

\section{Testing multiple experiments\label{sec:multi}}
Klaassen \cite{Klaassen2015}, see also \cite{Peeters2015}  suggested to obtain the evidential value $V$ for an article consisting of more than one experiment as the product of the evidential values $V_{j}$ of the single experiments in the article. The evidential value $V$ of a publication
given $N$ experiments is then
\begin{equation}
V=\prod_{j=1}^{N}V_{j}=\prod_{j=1}^{N}\underset{\vec{\rho}\in\mathcal{F}_{j}}{\max}\frac{f(z_{j}|\sigma_{Z}(\vec{\sigma}_{j},\vec{\rho}))}{f(z_{j}|\sigma_{Z}(\vec{\sigma}_{j},\vec{\rho}))}\,.
\end{equation}

Given that $V_{j}\ge1$, this immediately implies that the product grows exponentially with the number of experiments even if $H_{0}$ is true. Instead of obtaining the evidential value for every single experiment in an article, which (in a worst-case scenario) is based on a variance estimator from a single sample ($\sigma_{Z,j}^{2}=n_{j}z_{j}^{2}$), one may try to base that variance estimation on $N$ samples provided by the $N$ experiments in an article. I.e. 
\begin{equation}
V=\underset{\vec{\rho}\in\mathcal{F}}{\max}\prod_{j=1}^{N}\frac{f(z_{j}|\sigma_{Z}(\vec{\sigma}_{j},\vec{\rho}))}{f(z_{j}|\sigma_{Z}(\vec{\sigma}_{j},\vec{\rho}))}\,,
\end{equation}
where the feasible set $\mathcal{F}=\bigcap_{j=1}^{N}\mathcal{F}_{j}$, is just the intersect of all feasible sets $\mathcal{F}_{j}$ of every experiment. 

The idea of this alternative approach is simple: We cannot make a reliable statement about the probability of observing a single \emph{suspiciously} small $\tilde{z}_{j}$, particularly as $0=E[Z]$ under $H_{0}$. However, observing a \emph{suspiciously} small $\tilde{z}$ repeatedly is unlikely and may indicate sample correlations between groups. 

Following the \emph{worst-case} scenario above, the joint evidential value for $N$ experiments is asymptotically

\begin{eqnarray*}
\hat{V} & = & \underset{a\in(0,1]}{\max}a^{-N}\,\exp\left\{ -\sum_{j=0}^{N}\frac{n_{j}\, z_{j}^{2}}{2\, a^{2}\sigma_{0,j}^{2}}+\sum_{j=0}^{N}\frac{n_{j}\, z_{j}^{2}}{2\sigma_{0,j}^{2}}\right\} \\
 & = & \underset{a\in(0,1]}{\max}a^{-N}\,\exp\left\{ -\sum_{j=0}^{N}\frac{\tilde{z}_{j}^{2}}{2\, a^{2}}+\sum_{j=0}^{N}\frac{\tilde{z}_{j}^{2}}{2}\right\} \,,
\end{eqnarray*}
where again $\tilde{z}{}_{j}=\frac{\sqrt{n_{j}}z_{j}}{\sigma_{0,j}}$
and $\sigma_{0,j}=\sigma_{Z}(\sigma_{j},\vec{0})$. A straightforward computation reveals the surprisingly familiar result
\[
a^{2}=\frac{1}{N}\sum_{j=1}^{N}\tilde{z}^{2}\,.
\]

This implies that, in a \emph{worst-case} scenario, the joint likelihood-ratio compares a variance estimate based on $N$ samples with the expected one. And finally
\begin{eqnarray*}
\hat{V} & = & \begin{cases}
1 & :1\le\frac{1}{N}\sum_{j=1}^{N}\tilde{z}^{2}\\
\frac{\exp\left\{ -\frac{N}{2}+\frac{\sum_{j=0}^{N}\tilde{z}_{j}^{2}}{2}\right\}}{\left(\frac{1}{N}\sum_{j=1}^{N}\tilde{z}_{j}^{2}\right)^{\frac{N}{2}}}& :\,\text{else}.
\end{cases}
\end{eqnarray*}

Note that the joint evidential value for $N$ experiments relies on the fact that $\tilde{Z}_{j}\sim\mathcal{N}(0,1)$ i.i.d. under $H_{0}$ and therefore $\sum_{j=1}^{N}\tilde{Z}_{j}^{2}\sim\chi_{N}^{2}$. Hence the test statistics for sample correlations between groups can be expressed as a simple chi-squared statistic and one does not need to make the detour of obtaining an approximate distribution of $V$ under $H_{0}$.

\section{Relation to the $\Delta F$ test}
The $\chi^{2}$-test derived in the last section is closely related to the $\Delta F$-test suggested by the whistleblower \cite{Anonymous2012}. This test was also included in the report for the University of Amsterdam \cite{Peeters2015}.

Under $H_{0}$ and the assumption of a linear trend, the p-values of the $\Delta F$-test for a single experiment within an article are distributed uniformly in $[0,1]$. Using Fisher's method, it is then possible to obtain a p-value for an article comprising several experiments. The major difference between these two methods is that the $\Delta F$-test first determines a p-value for every study and tests whether the resulting p-values $p_{j}$ are \emph{to good to be true} while the chi-square test introduced here assesses this value directly by inspecting whether the relative deviations form perfect linearity $\tilde{z}_{j}^{2}$ are \emph{to good to be true}. Therefore, unsurprisingly, the two methods yield very similar results (see Table \ref{tab:pvalues}).

\begin{table}[!ht]
 \centering
 \begin{tabular}{|c|c|c|c|} \hline 
  Article & $\chi^{2}$-test & $\Delta F$-tests & Classification \tabularnewline \hline \hline 
  JF09.JEPG \cite{Forster2009a} & 8.06e-07 & 2.30e-07 & strong\tabularnewline \hline 
  JF11.JEPG \cite{Forster2011} & 8.73e-07 & 3.53e-07 & strong\tabularnewline \hline 
  JF.D12.SPPS \cite{Vision2014} & 7.14e-09 & 1.82e-08 & strong\tabularnewline \hline \hline
  L.JF09.JPSP \cite{Liberman2009} & 6.44e-4 & 8.46e-5 & strong\tabularnewline \hline
  L.JF09.JPSP* & 0.03 & 0.02 & -- \tabularnewline \hline \hline
  JF.LS09.JEPG \cite{Forster2009} & 0.25 & 0.11 & strong\tabularnewline \hline 
  JF.LK08.JPSP \cite{Forster2008} & 0.81 & 0.66 & inconclusive\tabularnewline \hline 
  D.JF.L09.JESP \cite{Denzler2009} & 0.93 & 0.52 & inconclusive\tabularnewline \hline 
  Reference \cite{Hagtvedt2011, Hunt2008, Kanten2011, Lerouge2009, Malkoc2010, Polman2011, Rook2011, Smith2008, Smith2006} & 0.11 & 0.14 & -- \tabularnewline \hline 
 \end{tabular}
 \caption{Comparison of p-values obtained with the direct $\chi^{2}$ and $\Delta F$ tests for studies classified as providing strong or inconclusive statistical evidence for low veracity by Peeters et al. \cite{Peeters2015}. The first three studies listed in the table were reported by the whistleblower \cite{Anonymous2012}. Note the divergence for JF.LS09.JEPG between the present analysis and \cite{Peeters2015}. Only those studies from \cite{Peeters2015} were considered here which provide at least $8$ experiments.  \label{tab:pvalues}}
\end{table}

\begin{figure}[!ht]
 \centering
 \includegraphics[width=0.75\textwidth]{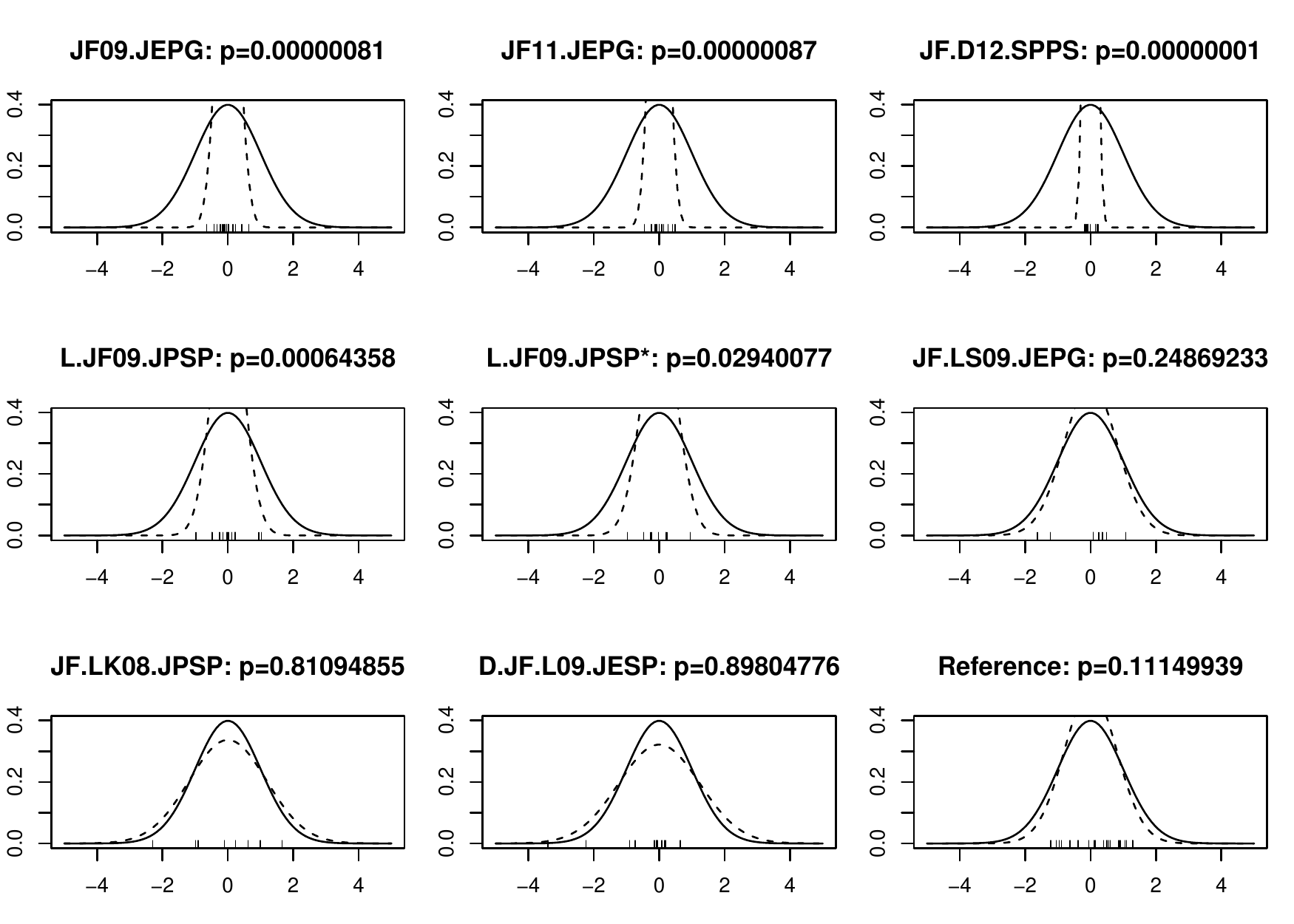}
 \caption{The distribution of $\tilde{z}_j$ (short dashes at the bottom of each panel) for each experiment from the articles listed in Table \ref{tab:pvalues}. The solid line shows the expected distribution of $\tilde{Z}_j$ under $H_0$ while the dashed line shows the normal distribution with $0$-mean and the variance estimated from the samples $\tilde{z}_j$. \label{fig:ztilde}}
\end{figure}

\begin{figure}[!ht]
 \centering
 \includegraphics[width=.75\textwidth]{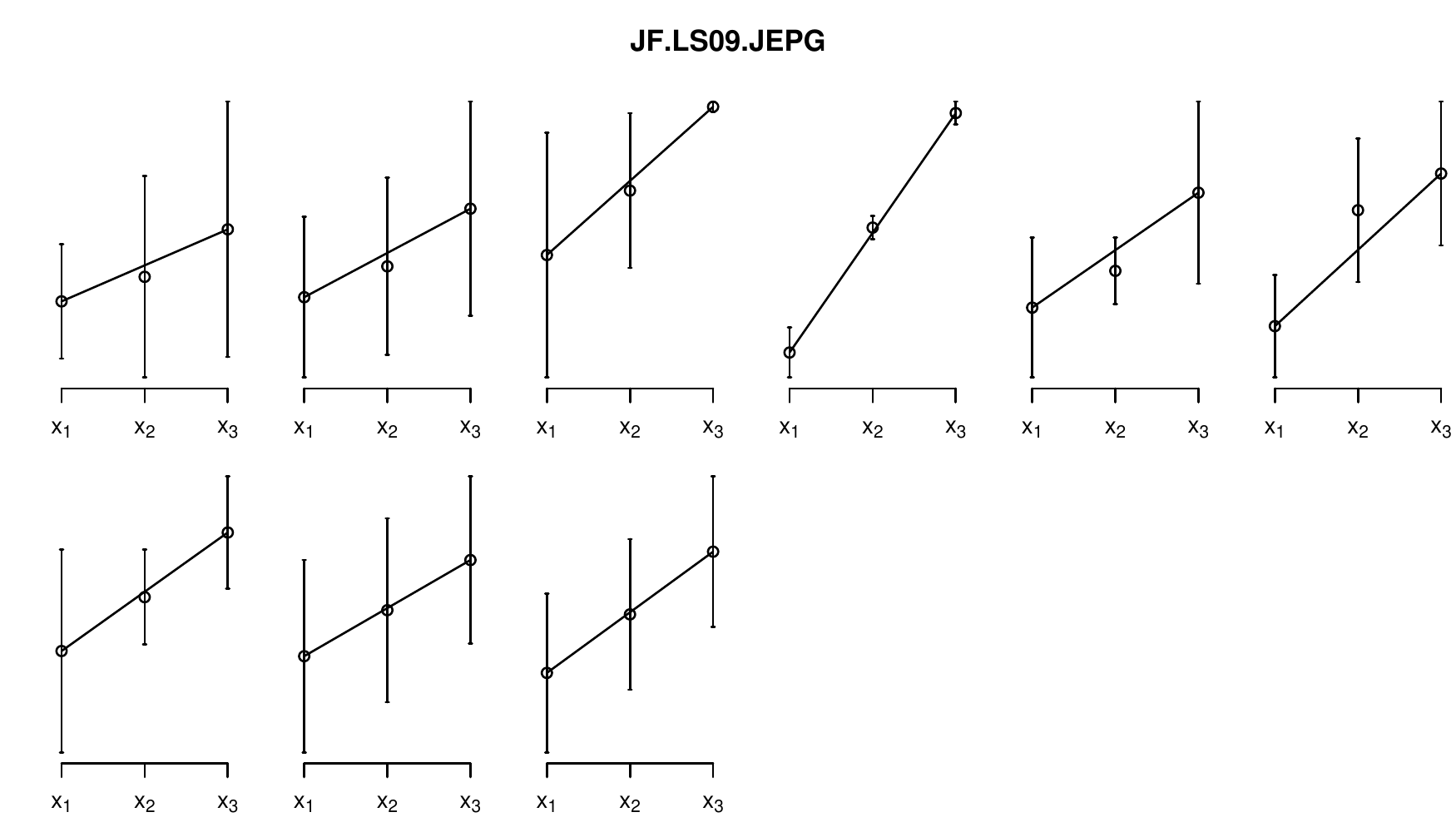}
 \caption{Condition means and stdandard deviations for 9 experiments from \cite{Forster2009}. \label{fig:jfls09}}
\end{figure}

Both methods, the $\chi^2$ and $\Delta F$ tests, are \emph{conservative} compared to the V-value approach by Klaassen \cite{Klaassen2015}. For example, the article {JF.LS09.JEPG} in Table \ref{tab:pvalues} was classified with \emph{strong statistical evidence for low veracity} \cite{Peeters2015} (compare also Figure \ref{fig:jfls09}). In contrasts, the $\chi^2$ and $\Delta F$ methods, yield p-values of $\approx 0.25$ and $\approx 0.11$, respectively, suggesting that there is no evidence of sample correlations between groups. The three methods agree for the studies {JF.LK08.JPSP} and D.JF.L09.JESP which were classified with \emph{inconclusive statistical evidence for low veracity}. The three methods also agree on classifying the three articles reported by the  whistleblower \cite{Anonymous2012} with \emph{strong statistical evidence for low veracity}. 

Depending on the chosen level of significance, the article L.JF09.JPSP could be classified as \emph{strong} or \emph{inconclusive}. This article contains conditions for which the authors did not expected a specific rank ordering of the condition means. Peeters et al. \cite{Peeters2015} included these \emph{control conditions} but reordered them according to increasing group means, yielding a p-value for the $\chi^2$-test of about $0.0006$ (L.JS09.JPSP in Table \ref{tab:pvalues}). Although the assumption of equidistant group means, i.e. $0=\mu_1-2\mu_2+\mu_3$, contains the assumption of equal group-means, i.e. $\mu_1 = \mu_2 = \mu_3$ as a special case, the actual test-result depends on the ordering of the conditions. Keeping the order of conditions as reported in \cite{Liberman2009} yields a p-value of about $0.015$ and excluding them results in a p-value of about $0.03$, shown as L.JF09.JPSP* in Table \ref{tab:pvalues}.

The discrepancy between the $\chi^2$ or $\Delta F$ methods and the V-value method for the JF.LS09.JEPG article \cite{Forster2009} is due to the tendency of the V-value method to indicate \emph{strong evidence} if a single experiment out of a series of experiments has a very small $\tilde{z}$-value. In contrast to the V-value method, the $\chi^2$ and the augmented V-method (see Section \ref{sec:multi}) take all experiments of an article into account by assuming the same correlation structure for all experiments. 

For the particular article \cite{Forster2009}, the V-value approach reported \emph{strong evidence for low  veracity} because the last two experiments (compare Figure \ref{fig:jfls09}) exhibit the \emph{super linear} pattern associated with sample correlations. The $\chi^2$ and $\Delta F$ method, however, do not indicate significant sample correlations as the deviance of remaining experiments fit well into the expected distribution under $H_0$, especially the results in panels 5 \& 6 in Figure \ref{fig:jfls09}.

Klaassen \cite{Klaassen2015} intended the V-value to be sensitive for single experiments. The argument is  that \emph{bad science cannot be compensated by very good science} \cite{Klaassen2015}. Finding a small value for $\tilde{z}_j$ in a series of experiments, however, is quiet probable\footnote{I.e. for $10$ experiments ($N=10$) $p\approx 0.4$ for $\alpha=0.05$ and $p\approx 0.1$ for $\alpha=0.01$} even under $H_0$. Hence one could argue that a single \emph{suspiciously small} $\tilde{z}_j$ can not be interpreted as strong evidence for sample correlations.

\section{Discussion}
There is no doubt that, in principle, statistics can be used to detect sample correlations that are due to data manipulation. The approach proposed in \cite{Klaassen2015}, however, is not without problems.

A first problem is the missing test statistics for the evidential value $V$. Although an upper-bound asymptotic test statistics for the V-value of a single experiment can be obtained (see Section \ref{sec:teststat} above and \cite{Klaassen2015}), the reliability of the $V$ value for a small $n$ remains unknown (as well as how large a large $n$ must be to be considered \emph{large}).

A second problem is the critical value of $V^{*}=6$ chosen by the authors,  which implies (asymptotically) $p\approx0.08$. Arguably, this is a rather high probability of falsely accusing a colleague of data manipulation.

A third problem is the assumption that the product of the evidence provided by every single experiment in an article can serve as a metric of evidence for data manipulation in this article. As mentioned above as well as in the comments to the article at pubpeer.com \cite{Pubpeer2015} and in a response by Denzler and Liberman \cite{Liberman2015}, this assumption implies that the evidence for data manipulation grows exponentially with the number of experiments even under $H_{0}$. The probability of $V\ge2$ for a single experiment is about $p\approx0.25$. Thus, about every 4th \emph{good} experiment will double the evidence for data manipulation. 

The fourth problem, finally, is a general concern. The analysis assumes a specific type of data manipulation. If this is true, the manipulation will induce correlations between condition means. Moreover, under the second assumption that $0=X_{1}-2X_{2}+X_{3}$ this correlation can be detected. Importantly, however, the reverse is not  true: The detection of such correlations in the data does not necessarily imply that data were manipulated. For that reason, Peeters et al. carefully avoided in \cite{Peeters2015} to claim that their findings prove that data were manipulated. Instead the  results are interpreted as \emph{evidence for low data veracity}, which is justified. In \cite{Klaassen2015}, however, Klaassen claims that its method provides evidence for manipulation. Although the origin of sample correlations cannot be determined with statistics, their presence certainly violates an ANOVA assumption. This may result in an increased type-I error rate. Therefore, the effects reported in the articles providing strong or possibly even inconclusive evidence for sample correlations (e.g \cite{Forster2009a,Forster2011,Vision2014,Liberman2009}) may be less significant than suggested by their ANOVAs. 

In this comment, specifically in Section \ref{sec:multi}, the concept of the single-experiment evidential value  was extended to multiple experiments. Moreover, a much simpler chi-squared test was provided to test the presence of correlations in the data that is similar to the test proposed in \cite{Anonymous2012} and yielded very similar probabilities for the presence of sample correlations. Thus, the V-value approach can serve as a test for sample correlations, if it is applied across several identical or at least similar experiments. In this case one is also able to decide whether the variability in the results is suspiciously small or not. However, estimating $\sigma_{Z}$ on the basis of a single experiment will certainly not reveal a reliable result.

\bibliographystyle{plain}
\bibliography{references}

\end{document}